\documentclass[aps,prb,twocolumn,showpacs,amsmath,amssymb,superscriptaddress,floatfix,citeautoscript]{revtex4-1}

\usepackage{graphicx}
\usepackage{dcolumn}
\usepackage{bm}
\usepackage{amssymb}
\usepackage{booktabs}
\usepackage{color}
\usepackage{microtype,lmodern}
\usepackage[T1]{fontenc}

\usepackage{sidecap}
\renewcommand{\figurename}{Fig.}
\renewcommand{\tablename}{Table}
\makeatletter\renewcommand{\fnum@figure}[1]{\figurename~\thefigure~(color online).}\makeatother
\makeatletter\renewcommand{\fnum@table}[1]{\tablename~\thetable.}\makeatother

\renewcommand{\Re}{\operatorname{Re}}
\renewcommand{\Im}{\operatorname{Im}}

\usepackage[colorlinks,plainpages=false,linkcolor=blue,urlcolor=blue,citecolor=blue,pdfpagemode=UseNone,pdfstartview=FitBH]{hyperref}

\begin{document}

\title{Quantitative determination of bond order and lattice distortions in nickel oxide heterostructures by resonant x-ray scattering}

\author{Y.~Lu}
\affiliation{Max-Planck-Institut f\"ur Festk\"orperforschung, Heisenbergstrasse~1, D-70569 Stuttgart, Germany}

\author{A.~Frano}
\affiliation{Max-Planck-Institut f\"ur Festk\"orperforschung, Heisenbergstrasse~1, D-70569 Stuttgart, Germany}
\affiliation{Helmholtz-Zentrum Berlin f\"ur Materialien und Energie, Wilhelm-Conrad-R\"ontgen-Campus BESSY II, Albert-Einstein-Strasse 15, D-12489 Berlin, Germany}

\author{M.~Bluschke}
\affiliation{Max-Planck-Institut f\"ur Festk\"orperforschung, Heisenbergstrasse~1, D-70569 Stuttgart, Germany}
\affiliation{Helmholtz-Zentrum Berlin f\"ur Materialien und Energie, Wilhelm-Conrad-R\"ontgen-Campus BESSY II, Albert-Einstein-Strasse 15, D-12489 Berlin, Germany}

\author{M.~Hepting}
\affiliation{Max-Planck-Institut f\"ur Festk\"orperforschung, Heisenbergstrasse~1, D-70569 Stuttgart, Germany}

\author{S.~Macke}
\affiliation{Max-Planck-Institut f\"ur Festk\"orperforschung, Heisenbergstrasse~1, D-70569 Stuttgart, Germany}

\author{J.~Strempfer}
\affiliation{Deutsches Elektronen-Synchrotron DESY, Notkestrasse~85, D-22607~Hamburg, Germany}

\author{P.~Wochner}
\affiliation{Max-Planck-Institut f\"ur Festk\"orperforschung, Heisenbergstrasse~1, D-70569 Stuttgart, Germany}

\author{G.~Cristiani}
\affiliation{Max-Planck-Institut f\"ur Festk\"orperforschung, Heisenbergstrasse~1, D-70569 Stuttgart, Germany}

\author{G.~Logvenov}
\affiliation{Max-Planck-Institut f\"ur Festk\"orperforschung, Heisenbergstrasse~1, D-70569 Stuttgart, Germany}

\author{H.-U. Habermeier}
\affiliation{Max-Planck-Institut f\"ur Festk\"orperforschung, Heisenbergstrasse~1, D-70569 Stuttgart, Germany}

\author{M.~W.~Haverkort}
\affiliation{Max-Planck-Institut f\"ur Chemische Physik fester Stoffe, N\"othnitzer~Strasse~40, D-01187 Dresden, Germany}

\author{B.~Keimer}
\email{b.keimer@fkf.mpg.de}
\affiliation{Max-Planck-Institut f\"ur Festk\"orperforschung, Heisenbergstrasse~1, D-70569 Stuttgart, Germany}

\author{E.~Benckiser}
\email{e.benckiser@fkf.mpg.de}
\affiliation{Max-Planck-Institut f\"ur Festk\"orperforschung, Heisenbergstrasse~1, D-70569 Stuttgart, Germany}

\date{\today}

\begin{abstract}
We present a combined study of Ni \textit{K}-edge resonant x-ray scattering and density functional calculations to probe and distinguish electronically driven ordering and lattice distortions in nickelate heterostructures. We demonstrate that due to the low crystal symmetry, contributions from structural distortions can contribute significantly to the energy-dependent Bragg peak intensities of a bond-ordered NdNiO$_3$ reference film. For a LaNiO$_3$-LaAlO$_3$ superlattice that exhibits magnetic order, we establish a rigorous upper bound on the bond-order parameter. We thus conclusively confirm predictions of a dominant spin density wave order parameter in metallic nickelates with a quasi-two-dimensional electronic structure.
\end{abstract}

\maketitle

\section{Introduction}

The mechanisms underlying the formation and competition of collective order in correlated-electron systems are preeminent themes of current solid-state science. The relative stability of charge and spin order and their roles in driving unusual phenomena such as multiferroicity and superconductivity have recently been subjects of intense investigation in a diverse set of solids ranging from molecular crystals to metal oxides. Metal-oxide thin films and superlattices have emerged as a particularly fruitful research platform because of the potential for targeted control of the carrier density, dimensionality, and interaction strength of correlated-electron systems~\cite{Mannhart2010,Hwang2012}. To take full advantage of these opportunities, accurate measurements of the lattice structure and electronic order parameters as well as quantitative feedback between theory and experiment are essential.

\begin{figure*}[t]
  \includegraphics[width=\textwidth]{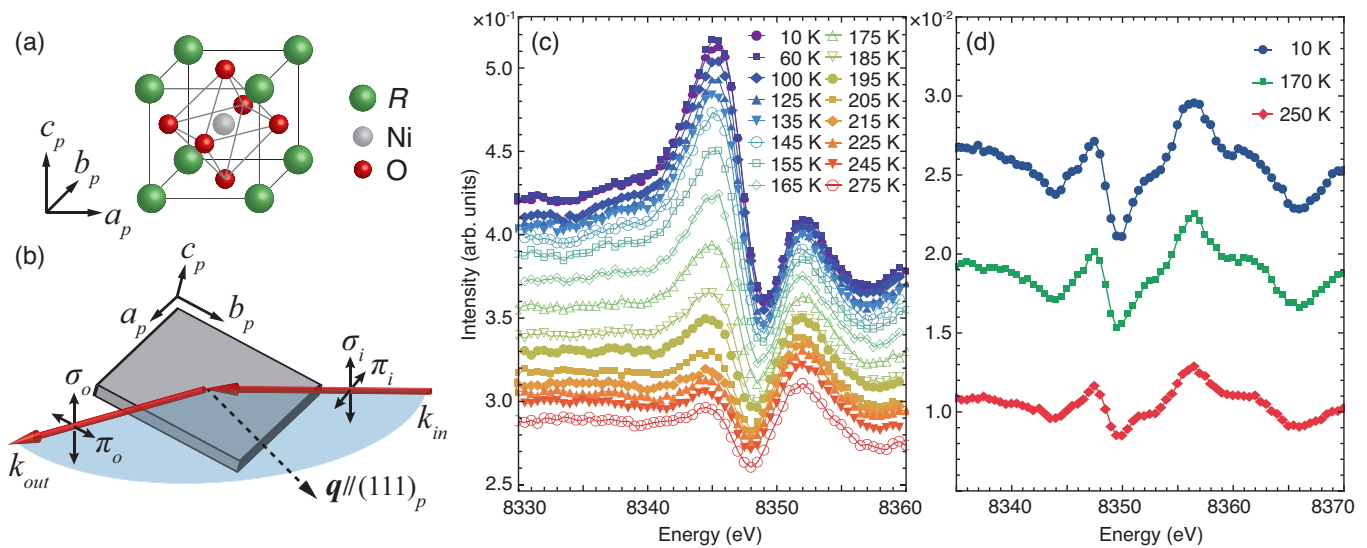}
  \caption{\label{fig1} Schematic of (a) a pseudocubic $R$NO unit cell and (b) the experimental geometry. The momentum transfer $\bm{q}=$ $\bm{k}_{out}-\bm{k}_{in}$ is parallel to the pseudocubic (111) direction. (c) and (d), Temperature dependence of the $(\frac{1}{2},\frac{1}{2},\frac{1}{2})_p$ reflection intensity of the (c) NNO thin film and (d) LNO-LAO SL around the Ni $K$ edge measured with $\sigma$-$\sigma$ polarization.}
\end{figure*}

A prototypical system of long-standing interest is the perovskite $R$NiO$_3$ ($R$ = rare-earth ion; $R$NO) whose approximately cubic unit cell is shown in Fig.~\ref{fig1}(a). In bulk $R$NO, the single-electron bandwidth is controlled by the Ni-O-Ni bond angle via tilts and rotations of the NiO$_6$ octahedra, which depend on the size of $R$. For the largest $R$ = La, the bonds are relatively straight, and $R$NO remains metallic and paramagnetic at all temperatures $T$. For smaller $R$, octahedral tilts and rotations reduce the bandwidth, and an insulating bond-ordered (BO) state with a two-sublattice array of Ni sites with long and short Ni-O bonds develops for $T \leq T_{BO}$. Antiferromagnetic spin order forms for $T \leq T_N$, with $T_N = T_{BO}$ for $R$ = Pr and Nd, and $T_N < T_{BO}$ for smaller $R$ with more highly distorted Ni-O-Ni bonds~\cite{Torrance1992,GarMun1994}. Various proposals have been made for the mechanism underlying the bond-ordering transition and its relation to magnetism. One set of models invokes charge order of Ni$^{2+}$ and Ni$^{4+}$ ions (electron configurations $3d^8$ and $3d^6$) that are stabilized by Hund's rule interactions relative to Ni$^{3+}$ ($3d^7$)~\cite{Mazin2007}. Alternative models attribute the bond disproportionation to a modulation of the Ni-O bond covalency without charge transfer between Ni sites~\cite{Quan2012,Johnston2014}. Finally, theories for the more itinerant $R$NO compounds indicate that bond order is secondary to magnetic order and predict a pure spin density wave (SDW) phase without charge or bond order for some lattice symmetries~\cite{Lee2011a,Lee2011b}. Recent experiments on $R$NO-based heterostructures with reduced dimensionality have provided some support for the latter prediction, but the bond-order parameter was probed only qualitatively through the formation of an optical gap~\cite{Boris2011,Jaramillo2014}, changes in the dc resistivity~\cite{Frano2013,Liu2013} and x-ray absorption spectra~\cite{Wu2015} at the metal-insulator transition, and characteristic Raman-active phonon modes~\cite{Hepting2014}. Quantitative tests of theoretical predictions have therefore not yet proven possible. This illustrates an important general challenge in the investigation of metal-oxide heterostructures, where the presence or absence of charge order---one of the most common order parameters in bulk metal oxides---is difficult to assess.

In the past two decades resonant x-ray scattering (RXS) has seen increasing use as a probe of charge, spin, and orbital ordering in 3$d$ transition metal oxides~\cite{Fink2013,Lorenzo2012,Matsumura2013,Beale2012}. At the transition metal $L$ edge, where a 2$p$ core electron is excited into the 3$d$ orbitals, one can study the states directly at the Fermi level. However, photons at this resonance have energies of a few hundred eV, which correspond to wavelengths too long to probe electronic order in many lattice structures. At the $K$ edge, on the other hand, the photon energy is an order of magnitude higher so that a larger portion of the reciprocal space can be reached. However, excitations at the $K$ edge probe the empty 4$p$ states of the transition metal, which are far above the Fermi level. These states are also affected by electronic ordering phenomena~\cite{Lorenzo2012,Matsumura2013,Beale2012,Joly2003,Balasubramanian2014,Akao2003}, but the coupling to the order parameter is less direct and hence more difficult to describe in a quantitative fashion~\cite{Nazarenko2006}. The influence of octahedral tilts and rotations further confounds the analysis and interpretation of $K$-edge RXS spectra~\cite{Takahashi2001,Takahashi2002}.

In this paper, we present a Ni $K$-edge RXS study of a LaNiO$_3$-LaAlO$_3$ superlattice (LNO-LAO SL) and a reference NdNiO$_3$ (NNO) film. The SL has been studied in previous work where a magnetically ordered ground state was revealed~\cite{Boris2011,Frano2013}. The complementary study we present here addresses the bond-order parameter that has been at the center of recent theoretical debate. We analyzed the RXS spectra using {\it ab initio} density functional theory (DFT) which accurately describes the weakly correlated $4p$ levels in the intermediate state of the RXS process in analogy to previous bulk studies~\cite{Nazarenko2006,Joly2003,Balasubramanian2014,Joly2001}. Aided by DFT calculations, both the bond-order parameter and the octahedral tilt and distortion pattern can be accurately extracted from the RXS data.

\section{Experimental and calculation details}

The high-quality LNO-LAO SL and NNO film studied were grown by pulsed laser deposition on (001)-oriented SrTiO$_3$ (STO) substrates with a lattice constant $a=3.905$~\AA\@. The LNO-LAO SL consists of 66 bilayers, each with two consecutive pseudocubic unit cells of LNO and equally thick LAO, corresponding to a total thickness of $\sim 100$~nm. The bilayer-averaged in- and out-of-plane lattice constants are $a_p=3.85$~\AA{} and $c_p=3.79$~\AA, respectively, as measured by hard x-ray diffraction. The $p$ denotes the pseudocubic perovskite unit cell [Fig.~\ref{fig1}(a)]. The NNO film is 40~nm thick with $a_p=3.88$~\AA{} and $c_p=3.77$~\AA\@. The room-temperature structural space groups of the NNO film and LNO-LAO SL were determined to be $Pbnm$ and $I2/c$, respectively, by analyzing a set of half-order reflections as described in Ref.~\onlinecite{Kinyanjui2014}. Previous studies showed that the magnetic order is robust against small structural variations and homogenously stabilized along the growth direction, although both samples show partial relaxation~\cite{Boris2011,Frano2013}.

The RXS measurements were performed at beamline P09 at PETRA III (DESY, Hamburg)~\cite{Strempfer2013} using a Si (111) monochromator with energy resolution of $\sim 1$~eV. A Cu (222) crystal was used for polarization analysis. Our RXS experiments focused on the pseudocubic $(\frac{1}{2},\frac{1}{2},\frac{1}{2})_p$ reflection [Fig.~\ref{fig1}(b)], which corresponds to the modulation wave vector of bond order in $R$NO.

DFT calculations were performed using the all-electron augmented plane waves plus local orbitals method implemented in the \textsc{wien2k} code~\cite{wien2k} with the Perdue-Burke-Ernzerhof~\cite{PBE} exchange-correlation functional on a $11 \times 8 \times 11$ $k$ mesh for NNO and on a $9 \times 3 \times 9$ $k$ mesh for LNO-LAO. The energy convergence was set to 0.1~mRy. The atomic positions are relaxed until the forces on each atom are smaller than 0.5~mRy/bohr.

\section{Results and discussion}

We first discuss the data on the NNO reference film. Figure~\ref{fig1}(c) shows the photon energy ($E$) dependence of the intensity of the $(\frac{1}{2},\frac{1}{2},\frac{1}{2})_p$ reflection in the vicinity of the Ni $K$-edge. The absorption edge $E_{\text{edge}}=$~8346~eV was defined as the maximum of the first derivative of the x-ray absorption spectrum. The intensities were collected with $\sigma$-$\sigma$ polarization at temperatures between 10 and 275~K during warming. A gradual decrease of the nonresonant pre-edge intensity is observed with increasing temperature, which can be explained by the Debye-Waller factor and a slight decrease of octahedral distortions~\cite{GarMun1992}. At 10~K, the energy dependence shows two maxima at 8345 and 8352~eV, similar to previously reported results which were interpreted as evidence of two distinct Ni valence states~\cite{Staub2002}. The amplitude of this two-peak structure decreases with increasing temperature and remains constant above $T_{\text{BO}}\sim 200~K$, qualitatively consistent with the melting of the bond order~\cite{Staub2002}. However, contrary to the observations in Ref.~\onlinecite{Staub2002} where the scattering intensity becomes independent of the photon energy above $T_{\text{BO}}$, the RXS intensity exhibits a nontrivial $E$ dependence up to room temperature.

To understand the remnant Ni resonant contribution above $T_{\text{BO}}$, we examine in detail the scattering intensity $I(\bm{q},E)=|F(\bm{q},E)|^2$, where the temperature term and Lorenz factor are neglected. The structure factor is given as $F(\bm{q},E)=\sum_j e^{\imath \bm{qr}_j}f_j(\bm{q},E)$, where $f_j(\bm{q},E) = f_j^{nr}(\bm{q}) + f_j^r (E)$ is the scattering factor of atom $j$ at position $\bm{r}_j$, and $\bm{q}$ denotes the x-ray momentum transfer. The summation runs over all atoms in the unit cell. $f_j^{nr}(\bm{q})=$ $(\bm{\varepsilon}^*_{o}\cdot \bm{\varepsilon}_{i}) f_j^0(\bm{q})$ is the nonresonant part of the atomic scattering factor, where $f_j^0(\bm{q})$ is the energy-independent Thomson scattering factor. $\bm{\varepsilon}_{i(o)}$ denotes the polarization of the incoming (outgoing) photons. The resonant contribution $f^r(E)=f'(E)+\imath f''(E)$ accounts for the energy dependence near the corresponding resonance edge, where $'$ and $''$ denote the real and imaginary part, respectively. In general, $f^r (E) \propto$ $\sum_{\alpha,\beta=1}^3 \varepsilon^*_{o,\alpha} \mathcal{F}_{\alpha\beta} \varepsilon_{i,\beta}$, where $\mathcal{F}$ is the scattering tensor at each $E$ with element
\begin{equation}
\label{ff1}
 \mathcal{F}_{\alpha\beta} = \sum_m\frac{\langle s |\hat T_\beta| p_m \rangle \langle p_m |\hat T_\alpha| s \rangle}{E-\delta E_m+\imath \frac{\Gamma}{2}}
\end{equation}
for the transition metal $K$ edge (1$s$$\rightarrow$4$p$) in the single particle approximation. $\hat T = (\hat x, \hat y, \hat z)$ is the dipole operator when considering only the dominant dipole-dipole contribution. $|s\rangle$ and $|p_m\rangle$ denote the 1$s$ and 4$p$ states, respectively, and $\delta E_m$ is their energy difference. $\Gamma$ is the core-hole lifetime of $\sim$1~eV~\cite{deGroot2008}. Considering the spherical symmetry of $|s\rangle$, one notices that $\Im \mathcal{F}$ is proportional to the 4$p$ density matrix $\mathcal{D}=|p\rangle \langle p|$ (with broadening) and
\begin{equation}
f''(E) \propto \sum_{\alpha,\beta=1}^3 \varepsilon^*_{o,\alpha} \varepsilon_{i,\beta} \mathcal{D}_{\alpha\beta}.
\end{equation}
Note that $\mathcal{D}$ is usually not diagonal, except for a few high symmetry cases.

\begin{figure}[tb]
  \includegraphics{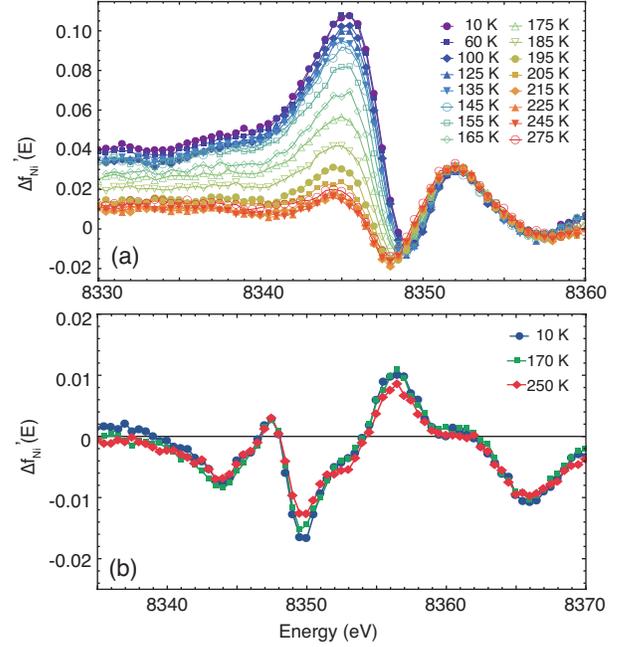}
  \caption{\label{fig2} Temperature dependence of $\Delta f_{Ni}'(E)$ for NNO thin film around the Ni $K$ edges.}
\end{figure}

For $R$NO, the structure factor of the $(\frac{1}{2},\frac{1}{2},\frac{1}{2})_p$ reflection is given as $F_{\frac{1}{2}\frac{1}{2}\frac{1}{2}}(\bm{q},E)=$ $A_{R\text{O}}(\bm{q}) + \Delta f_{\text{Ni}}^r(E)$, where $A_{R\text{O}}(\bm{q})$ is the non-resonant $R$ and O contribution. $\Delta f_{\text{Ni}}^r(E)=\sum_{j=1}^4 (-1)^j f_j^r(E)$ is the sum of resonant scattering factors of the four Ni sites in the orthorhombic or monoclinic unit cell. The Ni sites within one unit cell are not only related by a pure translation but with an additional rotation, resulting in oppositely-signed off-diagonal elements of $\mathcal{F}$ between different Ni sites, which give rise to a non-vanishing $\Delta f_{\text{Ni}}^r(E)$ and hence explain the remnant Ni resonant contribution observed in Fig.~\ref{fig1}(c). The real part of $\Delta f_{\text{Ni}}^r(E)$ can be approximately calculated by $\Delta f_{\text{Ni}}'(E)\approx$ $\sqrt{I(E)} - \sqrt{I(E_<)}$ from the measured spectra, with $E_<$ an off-resonance energy, assuming $A_{R\text{O}}'\gg $ $[A_{R\text{O}}''$, $\Delta f_{\text{Ni}}''(E)]$. The obtained result for $E_< = 8358$~eV is plotted in Fig.~\ref{fig2}(a). Note that in the bond-ordered phase, the slope of the spectra below $E_{\text{edge}}$ only decreases slowly with decreasing energy. This suggests that bond order modifies the RXS intensity in a wide range below $E_{\text{edge}}$ (more extended than the plotted range). In contrast, the spectral line shape above $E_{\text{edge}}$ is almost unaffected by the bond order, suggesting that it mainly originates from the octahedral tilts.

In previous RXS studies of the nickel oxides, the off-diagonal terms in $\mathcal{F}$ have been neglected, so that $\mathcal{F}$ could be effectively treated as a scalar for each Ni site~\cite{Staub2002,Scagnoli2005}. The resonant form factor $f^r(E)$ is then directly obtained by fitting both the energy-dependent RXS and x-ray absorption spectra, which are measures of the difference and sum of the form factors, respectively. In general, however, the contribution of off-diagonal terms in $\mathcal{F}$ can be significant and render the aforementioned fitting procedure inapplicable~\cite{Takahashi2001,Nazarenko2006}. In the NNO film, in particular, the RXS intensity is modified by the off-diagonal terms up to $\sim$20\% at room-temperature, as shown in Fig.~\ref{fig1}(c). The accurate description of the spectra and the quantitative determination of the bond-order parameter thus require detailed knowledge of the \textit{full} scattering tensor that cannot be obtained solely from experiment.

\begin{figure*}[tb]
  \includegraphics[width=\textwidth]{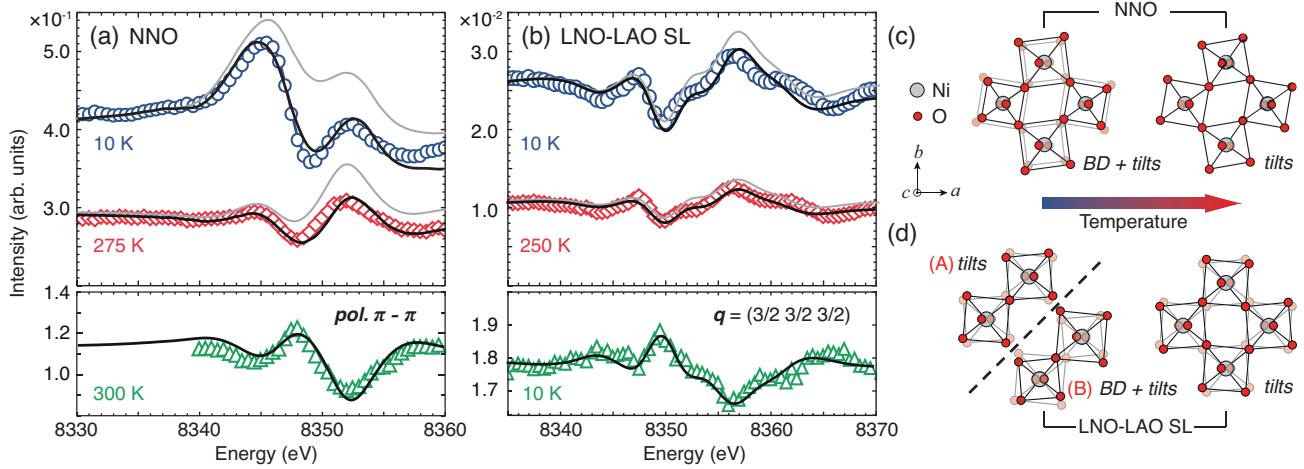}
  \caption{\label{fig3} (a) and (b) Measured (colored symbols) and calculated (black line) energy dependent intensity for (a) NNO thin film at the $(\frac{1}{2},\frac{1}{2},\frac{1}{2})_p$ reflection with $\sigma$-$\sigma$ polarization at 10 and 275~K (upper panel) or $\pi$-$\pi$ polarization at 300~K (lower panel), and for (b) LNO-LAO SL at the $(\frac{1}{2},\frac{1}{2},\frac{1}{2})_p$ reflection at 10 and 250~K (upper panel) or $(\frac{3}{2},\frac{3}{2},\frac{3}{2})_p$ at 10~K (lower panel). The calculated spectra without absorption correction are plotted as gray lines. (c) and (d) Top-view sketches of two consecutive NiO$_6$ octahedra layers in the (c) NNO film and (d) LNO-LAO SL at low (left) and room (right) temperatures. The bond disproportionation (BD) is exaggerated for better visualization.}
\end{figure*}

To this end, DFT calculations were performed to obtain atomic positions of the room-temperature structure within space group $Pbnm$ with the experimentally determined lattice constants. The low-temperature structure was obtained by DFT+$U$ relaxation in the magnetically ordered state with space group $P2_1/n$, using $U=5$~eV and $J=1$~eV. This approach has proven to yield structure parameters in good agreement with experimental results for the $R$NOs in both the metallic and the insulating states~\cite{May2010,Park2012}. The obtained atomic positions are summarized in the Appendix. The imaginary part of the scattering tensor, $\Im \mathcal{F}$, was obtained by scaling the DFT Ni 4$p$ density matrix to the tabulated Ni form factor~\cite{Chantler2000}. The real part, $\Re \mathcal{F}$, was calculated by Kramers-Kronig transformation. The $E$-dependent RXS spectra were finally calculated for the specific experimental geometry shown in Fig.~\ref{fig1}(b) following the formalism in Ref.~\onlinecite{Haverkort2010}. Considering the grazing incidence/emission required to reach the $(\frac{1}{2},\frac{1}{2},\frac{1}{2})_p$ reflection, the calculated spectra were multiplied by $A(E)=\frac{1}{2\mu(E)}(1-e^{-l \mu(E)})$ to account for the absorption effect, with $l$ the beam path length inside the sample at the corresponding reflection. Here $\mu(E)$ is the energy-dependent linear absorption coefficient and can be obtained by fitting the experimental absorption spectra to the tabulated values~\cite{Henke1993}.

\begin{figure}[b]
  \includegraphics{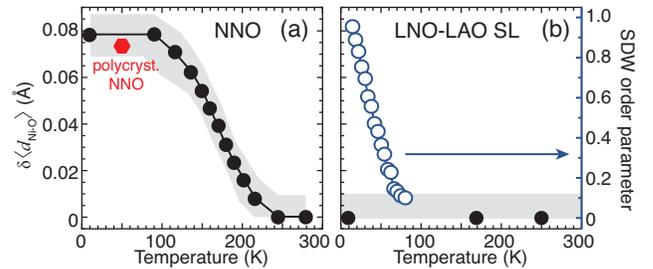}
  \caption{\label{fig4} (a) The fitted temperature dependence of $\delta\langle d_{\text{Ni-O}}\rangle$ in the NNO thin film. The experimentally determined value for a polycrystalline NNO~\cite{GarMun2009} is plotted for comparison. (b) Temperature dependence of $\delta\langle d_{\text{Ni-O}}\rangle$ in the LNO-LAO SL (solid symbols) and the SDW order parameter (open symbols)~\cite{Frano2013}. The gray areas indicate the error bars of the fitting.}
\end{figure}

Figure~\ref{fig3}(a) shows both the measured and calculated intensities of the $(\frac{1}{2},\frac{1}{2},\frac{1}{2})_p$ reflection at 10 and 275~K for the $\sigma$-$\sigma$ polarization channel, and at 300~K for $\pi$-$\pi$ polarization. The calculated spectra are scaled to match the measured nonresonant intensities at the corresponding temperatures. The drop of the nonresonant intensity across the Ni $K$ edge can be attrributed to the absorption effect, as illustrated by comparing the calculated spectra before and after the absorption correction. The calculated features in the energy dependence are in excellent agreement with the experimental data. We stress that the ratio of resonant and nonresonant intensities is also quantitatively reproduced, without adjustable parameters. The microscopic structural transition that leads to the RXS spectra line-shape change across $T_{\text{BO}}$ is depicted in Fig.~\ref{fig3}(c). Below $T_{\text{BO}}$, the Ni-O bond disproportionation between neighboring Ni sites that accompanies the bond order shifts the resonance energy and contributes mostly to the intensity modulation around 8345~eV. For the $P2_1/n$ structure, the calculated edge shift is 1.3~eV, slightly larger than the 1.2~eV determined for the polycrystalline bulk NNO~\cite{Medarde2009}. Upon warming, the bond disproportionation decreases while the tilt pattern of the octahedra remains the same. This is evidenced by the weakly $T$-dependent lineshape around 8352~eV.

\begin{figure}[b]
  \includegraphics{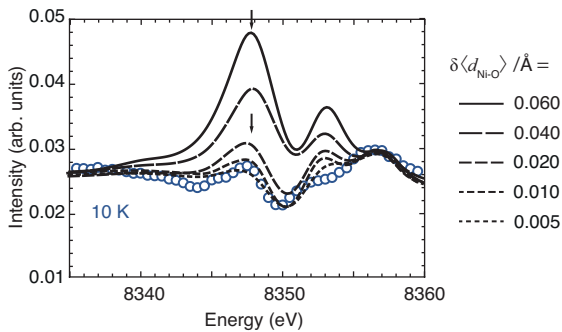}
  \caption{\label{fig5} Calculated energy dependent RXS intensity with $\sigma$-$\sigma$ polarization at $(\frac{1}{2},\frac{1}{2},\frac{1}{2})_p$ with $\delta\langle d_{\text{Ni-O}}\rangle=$ 0.005, 0.010, 0.020, 0.040, and 0.060~\AA.}
\end{figure}

The temperature dependence of the bond-order parameter, that is, the averaged Ni-O bond disproportionation $\delta\langle d_{\text{Ni-O}}\rangle$, can be obtained by fitting the $T$-dependent spectra using crystal structures interpolated between the DFT(+$U$) calculated $Pbnm$ and $P2_1/n$ unit cells. The results obtained in this way are plotted in Fig.~\ref{fig4}(a). The bond disproportionation at the lowest temperature is calculated as 0.077~\AA{}, in close agreement with the value determined for a polycrystalline NNO sample using x-ray powder diffraction~\cite{GarMun2009}. Upon warming, the bond disproportionation decreases in an order-parameter-like manner and vanishes above 200~K.

After demonstrating the quantitative accuracy of our methodology on the NNO reference sample, we now turn to the LNO-LAO SL\@ whose RXS spectra at $(\frac{1}{2},\frac{1}{2},\frac{1}{2})_p$ exhibit no noticeable temperature dependence except for an overall intensity decrease [Fig.~\ref{fig1}(d)]. The calculated $\Delta f_{\text{Ni}}^r(E)$ also remains unchanged within the experimental error [Fig.~\ref{fig2}(b)]. Figure~\ref{fig3}(b) shows that the spectra can be quantitatively reproduced by keeping all Ni sites in the same Wyckoff position, such that $\delta\langle d_{\text{Ni-O}}\rangle$ $=0$ [structure A in Fig.~\ref{fig3}(d)]. To estimate the effect of bond order in the observed RXS spectra, structure A was modified to incorporate various $\delta\langle d_{\text{Ni-O}}\rangle$ values while the rotation pattern of NiO$_6$ octahedra was kept the same [structure B in Fig.~\ref{fig3}(d)]. The calculated spectra shown in Fig.~\ref{fig5} were scaled such that the nonresonant intensity as well as the line shape above 8355~eV match the experimental spectrum. With increasing $\delta\langle d_{\text{Ni-O}}\rangle$, the intensity of the peak at $\sim$8346~eV (marked by arrows) increases, in analogy to the case of NNO. For $\delta\langle d_{\text{Ni-O}}\rangle$ $=0.01$\AA{} the calculation already deviates clearly from the experimental result. A rigorous upper bound of $\delta\langle d_{\text{Ni-O}}\rangle$ $< 0.01$~\AA{} [shaded area in Fig.~\ref{fig4}(b)] can thus be placed on the bond-order parameter. The absence of bond order together with the previously reported magnetic order below 100~K [Fig.~\ref{fig4}(b)] conclusively confirms the prediction of a pure SDW phase in the nickelate SLs~\cite{Lee2011a,*Lee2011b}.

\section{Conclusion}

In summary, we showed that RXS at the Ni $K$ edge in conjunction with \textit{ab initio} density functional theory is an accurate, quantitative probe of bond order in a NNO thin film and in a LNO-LAO superlattice. Whereas the bond-order parameter determined for NNO is in quantitative agreement with data on bulk NNO, the superlattice with atomically thin LNO layers shows no bond order at any temperature. This finding, together with the previously reported magnetic order below 100~K [Fig.~\ref{fig4}(b)] conclusively confirms the theoretical prediction of a pure SDW ground state in low-dimensional $R$NO systems~\cite{Lee2011a,Lee2011b}. In addition, our experiments revealed a significant Ni resonant contribution to the RXS spectra independent of bond order in both systems. With the aid of DFT calculations, we were able to attribute this contribution to octahedral distortions and tilts, and to extract highly specific information on the distortion pattern and amplitude from the experimental spectra. We emphasize that this information was obtained from measurements at a single Bragg reflection. The synergistic application of RXS and DFT therefore opens up unique perspectives for the characterization of electronic order and lattice symmetry in thin films and multilayers where only a few Bragg reflections are experimentally accessible due to geometric constraints.

\section*{Acknowledgments}

The RXS measurements were carried out at the light source PETRA III at DESY, a member of the Helmholtz- Gemeinschaft. We would like to thank Sonia Francoual for assistance at beamline P09. We thank Z.~Zhong, M.~H\"oppner, and D.~Kasinathan for helpful discussions. Financial support from the DFG Grant No. SFB/TRR80 project G1 is acknowledged. Y.~L. and A.~F. contributed equally to this work. 

\appendix*

\section{Calculated atomic positions and scattering tensors}

The structural parameters obtained by DFT(+$U$) for NNO and the LNO-LAO SL are summarized in Tables~\ref{tabA1} and \ref{tabA2}, respectively. The space group of LNO-LAO is $P2_1/c$ instead of the experimentally determined LNO-LAO averaged $I2/c$~\cite{Kinyanjui2014}, because of the additional symmetry breaking by substituting two NiO$_2$ layers with AlO$_2$ layers in the superlattice unit cell.

\begin{table}[t]
    \caption{\label{tabA1} DFT(+$U$) atomic positions for NNO with lattice parameters $a$=$b$=5.487\AA{}, $c$=7.546\AA.}
    \begin{tabular}{@{}clclclcl@{}} \toprule[0.08em]
       & & \phantom{aa} & x  & \phantom{aa} & y & \phantom{aa} & z  \\ \midrule[0.05em]
       &Nd     && 0.488 && 0.058 && 0.750 \\
$Pbnm$ &Ni && 0.000 && 0.000 && 0.500 \\
(DFT)  &O$_1$  && 0.591 && 0.477 && 0.750 \\
       &O$_2$  && 0.207 && 0.294 && 0.547 \\
      \midrule[0.05em]
       &Nd     && 0.489 && 0.056 && 0.750 \\
       &Ni$_1$ && 0.000 && 0.000 && 0.000 \\
$P2_1/n$  &Ni$_2$ && 0.000 && 0.000 && 0.500 \\
(DFT+$U$) &O$_1$  && 0.595 && 0.475 && 0.755 \\
       &O$_2$  && 0.198 && 0.291 && 0.549 \\
       &O$_3$  && 0.211 && 0.198 && 0.452 \\ \bottomrule[0.08em]
    \end{tabular}
\end{table}

\begin{table}[t]
    \caption{\label{tabA2} DFT atomic positions for LNO-LAO SL with lattice parameters $a$=$b$=5.438\AA{}, $c$=15.156\AA.}
    \begin{tabular}{@{}clclclcl@{}} \toprule[0.08em]
       & & \phantom{aa} & x  & \phantom{aa} & y & \phantom{aa} & z  \\ \midrule[0.05em]
       &La$_1$ && 0.997 && 0.500 && 0.875 \\
       &La$_2$ && 0.002 && 0.502 && 0.123 \\
       &La$_3$ && 0.998 && 0.500 && 0.375 \\
       &Ni && 0.501 && 0.500 && 0.000 \\
       &Al && 0.498 && 0.499 && 0.499 \\
$P2_1/c$ &O$_1$ && 0.438 && 0.500 && 0.875 \\
       &O$_2$ && 0.555 && 0.500 && 0.126 \\
       &O$_3$ && 0.451 && 0.500 && 0.375 \\
       &O$_4$ && 0.747 && 0.249 && 0.985 \\
       &O$_5$ && 0.253 && 0.753 && 0.015 \\
       &O$_6$ && 0.751 && 0.250 && 0.487 \\
       &O$_7$ && 0.249 && 0.749 && 0.513 \\ \bottomrule[0.08em]
    \end{tabular}
\end{table}

\begin{figure}[htb]
  \includegraphics{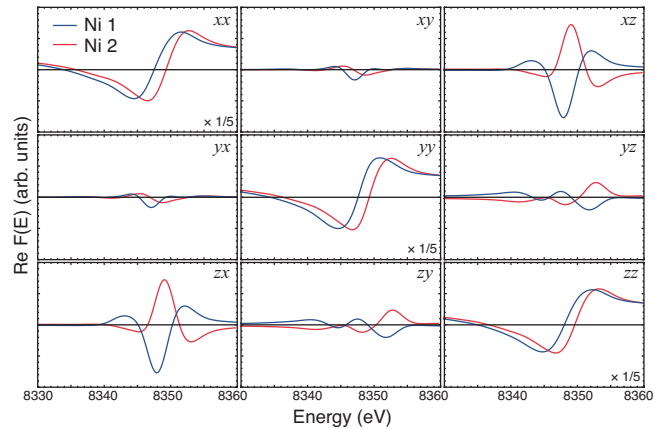}
  \caption{\label{figA1} The real part of the scattering tensor $\Re \mathcal{F}$ for two neighboring Ni sites in the $P2_1/n$ unit cell of NNO. The diagonal terms are shifted vertically and scaled by 1/5 for clarity.}
\end{figure}

The real part of the calculated scattering tensor for two neighboring Ni sites in the $P2_1/n$ unit cell of NNO is shown in Fig.~\ref{figA1} as an example. The column/row indices $x, y, z$ denote the local coordinates defined by the lattice vectors. The effect of bond order is reflected by an energy shift in the diagonal elements, and the octahedral distortions by inequivalent off-diagonal elements. For the $P2_1/n$ symmetry, both the shift in diagonal elements \textit{and} the oppositely signed $\mathcal F_{xz(zx)}$ are the leading contributions to the $(\frac{1}{2},\frac{1}{2},\frac{1}{2})_p$ reflection with $\sigma$-$\sigma$ ($\pi$-$\pi$) polarization.

\bibliographystyle{apsrev4-1}
\bibliography{RXS_Kedge}

\end{document}